\documentclass[prb,twocolumn,showpacs,floatfix,superscriptaddress]{revtex4-1}

\usepackage{graphicx}
\usepackage{array}
\usepackage{amsmath,amsfonts,amssymb}
\usepackage[ansinew]{inputenc}
\usepackage{color}
\usepackage{hyperref}

\newcommand{\Eqref}[1]{Eq.~(\ref{#1})}
\newcommand{\Figref}[1]{Fig.~\ref{#1}}
\newcommand{\Tabref}[1]{Table.~\ref{#1}}

\begin{document}

\title{Phonon-induced linewidths of graphene electronic states}

\author{Bo Hellsing}
\affiliation{\mbox{Donostia International Physics Center (DIPC) -- UPV/EHU,
E-20018 San Sebasti\'an, Spain}}
\affiliation{Material and Surface Theory Group, Department of Physics,
University of Gothenburg, Sweden}

\author{Thomas Frederiksen}
\affiliation{\mbox{Donostia International Physics Center (DIPC) -- UPV/EHU,
E-20018 San Sebasti\'an, Spain}}
\affiliation{IKERBASQUE, Basque Foundation for Science, E-48013, Bilbao, Spain}

\author{Federico Mazzola}
\affiliation{Department of Physics, Norwegian University of Science and Technology (NTNU), NO-7491 Trondheim, Norway}

\author{Thiagarajan Balasubramanian}
\affiliation{MAX-lab, PO Box 118, S-22100 Lund, Sweden}

\author{Justin W. Wells}
\affiliation{Center for Quantum Spintronics, Department of Physics, Norwegian University of Science and Technology (NTNU), NO-7491 Trondheim, Norway}

\begin{abstract}
The linewidths of the electronic bands originating from the electron-phonon coupling in graphene are analyzed based on model tight-binding calculations and experimental angle-resolved photoemission spectroscopy (ARPES) data. Our calculations confirm the prediction that the high-energy optical phonons provide the most essential contribution to the phonon-induced linewidth of the two upper occupied $\sigma$ bands near the $\bar{\Gamma}$-point. For larger binding energies of these  bands, as well as for the $\pi$ band, we find evidence for a substantial lifetime broadening from interband scattering $\pi \rightarrow \sigma$ and $\sigma \rightarrow \pi$, respectively, driven by the out-of-plane ZA acoustic phonons. The essential features of the calculated  $\sigma$ band linewidths are in agreement with recent published  ARPES data [F.~Mazzola \textit{et al.}, Phys.~Rev.~B. {\textbf 95}, 075430 (2017)] and of the $\pi$ band linewidth with ARPES data presented here.
\end{abstract}

\date{\today}


\maketitle

\section{Introduction}
Numerous experimental and theoretical studies of graphene have been presented during the last decade.\cite{CaGuPe.09.electronicpropertiesof,DaAdHw.11.Electronictransportin} These investigations have revealed remarkable mechanical \cite{LeWeKy.08.MeasurementElasticProperties}, electronic \cite{MoNoKa.08.GiantIntrinsicCarrier}, optical \cite{BoSuHa.10.Graphenephotonicsand}, and thermal \cite{BaGhBa.08.SuperiorThermalConductivity} properties. However, graphene is not considered to be a good BCS superconductor \cite{BaCoSc.57.TheorySuperconductivity} because of very weak electron-phonon coupling (EPC).\cite{SiLiDu.13.FirstPrinciplesCalculations} This is probably true when considering the $\pi$ bands crossing the Fermi level: for neutral graphene, two nearly-linearly dispersing bands touch at the Fermi level at the charge neutrality point, also called Dirac point. In this case, or even for extreme ranges of doping levels or electrostatic gating conditions, the density of states (DOS) is low, as well as the electron-phonon matrix element. On the other hand, the situation can be considerably different in other parts of the electronic spectrum. Recently, Mazzola \textit{et al.} \cite{MaWeYa.13.Kinksinsigma, Mazzola:2017} reported evidence of strong EPC in the $\sigma$-band; which was revealed in angle-resolved photoemission spectroscopy (ARPES) measurements with a substantial lifetime broadening and a pronounced kink in the dispersion.

The aim of the present combined theory and experimental study has been to get insight to the scattering process determining the EPC induced linewidths of the occupied $\sigma$ bands and $\pi$ band. Of particular interest was to investigate the relative importance of the \textit{intraband} and \textit{interband} scattering as well as which dominant phonon modes drive the scattering. To the best of our knowledge, linewidth analysis of the $\sigma$ bands is still missing in the literature. The phonon induced linewidth of the $\pi$ band has been studied by Park \textit{et al}. \cite{PaGiSp.09.FirstPrinciplesStudy} in the binding energy range 0-2.5 eV. In the lower part of this energy range our results agrees reasonably while in the upper part, our linewidths are about twice as large.

The \textit{intraband} scattering, which is found to be driven by the high energy in-plane optical phonons, is an important scattering channel for both the $\sigma$ and the $\pi$ band. For the two occupied uppermost $\sigma$ bands this channel dominates near the EPC induced ``kink'', about 200 meV below the top of these bands. However, the \textit{interband} $\pi \rightarrow \sigma$ and $\sigma \rightarrow \pi$ scattering can be mediated by the existence of out-of-plane vibrational modes. Our calculations reveal a substantial contribution from these scattering channels, driven by in particular the out-of-plane acoustic ZA mode, at higher binding energies.

The paper is organized as follows. In next section, Sec.~II we introduce the theoretical formulation of the EPC linewidth and outline the calculation of the electron and phonon band structure. In addition we give some details about the approximations used when constructing the deformation potential. In Sec.~III we present the results of the linewidth calculations for $\sigma$ bands and the $\pi$ band and compare with experimental data.
Our summary and conclusions and some perspectives for future research are presented in Sec.~IV.

\section{EPC-induced linewidth}
Our calculations are based on the traditional theoretical framework where the distortion of the electronic Hamiltonian caused by lattice vibrations can be considered to be of first order. 

In the low-temperature limit, which is the relevant case in the experiment reported by Mazzola \textit{et al}. \cite{MaWeYa.13.Kinksinsigma}, the thermal energy ($\approx 6$~meV) is less than the typical phonon energy ($\approx 170$~meV). In this case phonon emission dominates, while phonon absorption is suppressed. 

The EPC contribution to the linewidth of a particular electron band $n$ and wave vector $\mathbf k$ is calculated applying first order time dependent perturbation theory, the Fermi Golden Rule
\begin{eqnarray}
&&\Gamma_{n \mathbf k} =  \\ 
 && 2 \pi\sum_{n' \nu \mathbf q}|\langle n \mathbf k  |\delta V^{\nu}_{\mathbf q}|n' \mathbf k + \mathbf q\rangle |^2 \delta(\varepsilon_{n'\mathbf k + \mathbf q}-\varepsilon_{n \mathbf k } - \hbar \omega_{\nu \mathbf q}) \ , \nonumber 
\label{eq:gamma}
\end{eqnarray}
where $\varepsilon_{n \mathbf k }$ and $\omega_{\nu \mathbf q}$ represent the electron band energy and phonon frequency, respectively. The phonons
are described by band index $\nu$ and wave vector $\mathbf{q}$.
In the harmonic approximation the deformation potential is written
\begin{eqnarray}
&&\delta V_{\mathbf q}^{\nu}(\mathbf r) = \\ && \sqrt{\frac{\hbar}{2M\omega_{\nu\mathbf q}}}\sum_{\mathbf R}  \mathbf e^{\nu}(\mathbf q)\cdot \mathbf V'(\mathbf R + \mathbf r_s; \mathbf r) e^{-i\mathbf q \cdot (\mathbf R + \mathbf r_s)} \nonumber \ , 
\label{eq:phonon}
\end{eqnarray}
where $\mathbf{R}$ denotes the center position of the unit cells and $\mathbf{r_s}$ the positions of the A and B atoms within the unit cell. $\mathbf e^{\nu}(\mathbf q)$ is a six dimensional polarization vector with components $e^{\nu}_{si}(\mathbf q)$, where $s$=(A,B) and index $i$ refers to the three Cartesian coordinates of the displacement vector \( \mathbf{X}=(X,Y,Z) \). The derivative of the one-electron potential $\mathbf V'$ has six components $V'_{si}= {\partial V_s}/{\partial X_i} $. 

A calculation of the EPS linewidth apparently requires information about the electron structure -- band structure and wave functions, and phonon structure -- band structure and polarizations fields. The electron structure is achieved from a tight-binding (TB) calculations and the phonon structure from a force constant model (FCM).


\subsection{Electron structure}

\begin{figure}
\includegraphics[width=0.9\columnwidth]{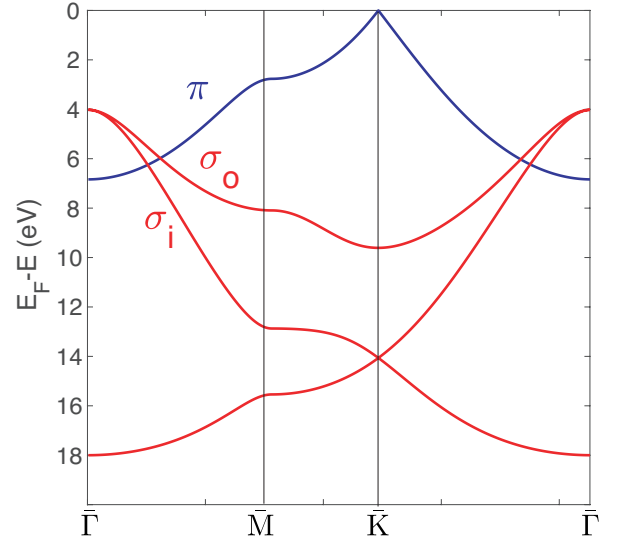}
\caption{(Color online) Occupied part of the electron band structure. The $\sigma$ bands in red and the $\pi$ band in blue.}
\label{ebands}
\end{figure}

In the TB approximation the wave functions $\psi_{n\mathbf k}$ are written  
 \begin{eqnarray}
  \psi_{n\mathbf k}(\mathbf r) = \sum_{js} c_{nsj}(\mathbf k) \Psi_{sj}(\mathbf k,\mathbf r) \ , 
\label{wave} 
\end{eqnarray}
where the Bloch orbitals are given by
 \begin{eqnarray}
  \Psi_{sj}(\mathbf k,\mathbf r) = \frac{1}{\sqrt{N}} \sum_{\mathbf R} \phi_{j}(\mathbf r - (\mathbf R + \mathbf r_s )) e^{i\mathbf k\cdot(\mathbf R+\mathbf r_s)} ,
 \end{eqnarray}
%
%
where $N$ denotes the number of unit cells to be summed over, $\phi_j$ the basis $\{\phi_{2s},\phi_{2p_x},\phi_{2p_y},\phi_{2p_z}\}$. 
The electronic bands $\varepsilon_{n\mathbf k}$ and coefficients $c_{nsj}(\mathbf k)$ are obtained by solving the generalized eigenvalues problem:
\begin{eqnarray}
 \sum_{J'} [H_{JJ'}(\mathbf k) - \varepsilon_{n\mathbf k} S_{JJ'}(\mathbf k)]c_{nJ'}(\mathbf k) = 0 \ ,
\end{eqnarray}
where we use the short hand index notation $J=js$. $S_{JJ'}$ denotes the overlap matrix elements. We apply the TB parameter-set shown in \Tabref{table_el}.
\begin{table}
\vspace{-0.3cm}
\caption{Tight binding parameters. Direct terms $\varepsilon_{2s}$ and $\varepsilon_{2p}$ and hopping parameters $V_{ss\sigma}$, $V_{sp\sigma}$, $V_{pp\sigma}$ and $V_{pp\pi}$ are all given in units of eV while the overlap parameters $S_{ss\sigma}$, $S_{sp\sigma}$, $S_{pp\sigma}$ and $S_{pp\pi}$ are dimensionless. Based on published values of these parameters, \cite{SaDrDr.98.PhysicalPropertiesCarbon,Gharekhanlou:2011,KoGmFa.10.Tightbindingtheory} we have made some slight adjustments to fit our own previously published DFT band structure calculation. \cite{Mazzola:2017}}
\centering
\begin{tabular}{ c c c c c c } 
\hline\hline
$\varepsilon_{2s}$  & $\varepsilon_{2p}$ & $V_{ss\sigma}$  & $V_{sp\sigma}$ & $V_{pp\sigma}$  & $V_{pp\pi}$ \\  
$-8.70$  & 0.00 & $-6.70$ & 5.50  & 5.90 &  $-3.10$ \\
\hline 
& $S_{ss\sigma}$ & $S_{sp\sigma}$ & $S_{pp\sigma}$ & $S_{pp\pi}$ \\  
& 0.20 & $-0.10$ & $-0.15$ & $0.12$\\
\hline\hline
\end{tabular}
\label{table_el} 
\end{table}
The calculated band structure is shown in  \Figref{ebands} and agrees well with our Density functional theory (DFT) based calculation published recently.\cite{Mazzola:2017}

\subsection{Phonon structure}

All six phonon modes are considered, three optical and three acoustic.
The optical phonon modes are: longitudinal optical (LO), transversal optical (TO) and out-of-plane optical (ZO).
The acoustic phonon modes are: longitudinal acoustic (LA), transversal acoustic (TA) and the out-of-plane acoustic (ZA). Applying a FCM the dynamical matrix $D$ is calculated including up to third order nearest neighbor interactions. The force constants $\Phi_{ll'}^{As}$ are defined by
\begin{eqnarray}
  D_{ll'}^{As}(\mathbf q) = \sum_{\mathbf{R}_s} \Phi_{ll'}^{As}(\mathbf{R}_s)e^{-i\mathbf q\cdot\mathbf{R}_s} ,
  \label{dyn}
\end{eqnarray}
where index $l$ denotes the components of a complex vector $(\xi,\eta)$ where $\xi=X+iY$ and $\eta=X-iY$ and $\mathbf{X}_{\parallel}=X\hat{x}+Y\hat{y}$ being the atomic in-plane displacement vector. $\mathbf{R}_s$ labels the vectors from a center A atom to the three nearest B atom, the six next-nearest A atoms and the three next-next-nearest B atoms. 

The in-plane force constants, in the $(\xi,\eta)$ representation, are parametrized according to Falkovsky.\cite{Falkovsky:2008}
To achieve the dynamical matrix elements in the $(X,Y)$ representation we have to transform the force constants in the $(\xi,\eta)$ representation to the $(X,Y)$ representation. We then derive
\begin{eqnarray}
D^{As'}_{XX}(\mathbf q) &=& 2D^{As'}_{\xi\eta}(\mathbf q) + D^{As'}_{\xi\xi}(\mathbf q) + D^{As'}_{\eta\eta}(\mathbf q) \nonumber \\
D^{As'}_{YY}(\mathbf q) &=& 2D^{As'}_{\xi\eta}(\mathbf q) - D^{As'}_{\xi\xi}(\mathbf q) - D^{As'}_{\eta\eta}(\mathbf q) \nonumber \\
D^{As'}_{XY}(\mathbf q) &=& i[D^{As'}_{\xi\xi}(\mathbf q) - D^{As'}_{\eta\eta}(\mathbf q)] \ .
\label{repr4}
\end{eqnarray}
Fourier transforming the equation of motion we then get the eigenvalue problem:
\begin{eqnarray}
 \sum_{s'i'} [D_{ii'}^{ss'}(\mathbf q) - \omega^2_{\nu}(\mathbf{q}) \delta_{ss'}\delta_{ii'}]e^{\nu}_{s'i'}(\mathbf q) = 0 \ ,
\label{eigen}
\end{eqnarray}
\begin{table}
\vspace{-0.3cm}
\caption{Force constants in units of 10$^5$ cm$^{-2}$ for nearest-neighbors (nn), next-nearest-neighbors (nnn) and next-next-nearest-neighbors (nnnn)  interaction. The force constants are given in the complex representation $(\xi,\eta)$, where $\xi=X+iY$ and $\eta = X - iY$, where $(X,Y)$ is the Cartesian coordinate representation. 
All force constants are taken from Falkovsky \cite{Falkovsky:2008}, except  $\Phi^{SUB}_{zz}$ which is the force constant representing a spring connecting the carbon atoms to a rigid substrate. The value 0.38 for this force constant is set to reproduce the finite energy of the ZA mode at $\bar{\Gamma}$ of 24 meV in a recent DFT based calculation of graphene on SiC.\cite{Minamitani:2017}} 
\centering
\begin{tabular}{| c | c | c | c | c | c |} 
 \hline\hline
 &
 \\
nn \ &  \ \  $\Phi^{AB}_{\xi\eta}$ \ \ \ \ \ $\Phi^{AB}_{\xi\xi}$  \ \ \ \ \  $\Phi^{AB}_{zz}$ \ \ \ \ \ \ $\Phi^{SUB}_{zz}$  \\ 
 &
 \\
 & \ \ \ \ -4.095  \  \ \ \ -1.645   \ \ \ -1.415  \ \ \ \ 0.380 \ \ \ \  
\\
 \hline 
 &
 \\
nnn \ &   \ $\Phi^{AA}_{\xi\eta}$  \ \ \ \ \ $\Phi^{AA}_{\xi\xi}$  \ \ \ \ \ $\Phi^{AA}_{zz}$  \\  
&
\\
& 
\ \ -0.209 \  \ \ \ \  0.690 \ \ \ 0.171 \ \ \    \\
 \hline 
&
\\
 \ nnnn \ &  \ $\Phi^{AB}_{\xi\eta}$   \ \ \ \ \ $\Phi^{AB}_{\xi\xi}$ \ \ \ \ \  $\Phi^{AB}_{zz}$  \\
&
\\
& 
\ \ -0.072  \ \ \ \ \    0.375   \  \ \ \ \  0.085 \ \ \    \\
 \hline
\end{tabular}
\label{table_ph} 
\end{table}
where the subscript $i$ labels the three components $X$, $Y$ and $Z$ of the Cartesian displacement vector  $\mathbf{X}=X\hat{x}+Y\hat{y}+Z\hat{z}$, where $Z$ denotes the out-of-plane displacement. 

In addition to the parameters of Falkovsky \cite{Falkovsky:2008}, we add a spring between all carbon atoms of the graphene layer and a rigid substrate in order to take into account the influence of the substrate in a first order approximation. The force constant of this spring is adjusted to fit the out-of-plane phonon mode dispersion from a $\it{first}$ $\it{principles}$ calculation of the  phonon band structure of graphene on SiC.\cite{Minamitani:2017} 

\begin{figure}
\includegraphics[width=0.95\columnwidth]{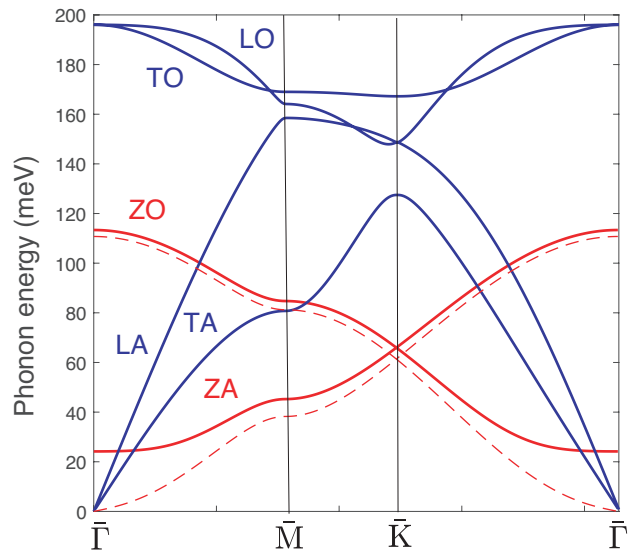}
\caption{(Color online) Phonon band structure of unsupported graphene (dashed lines) and graphene supported on a rigid substrate (solid lines). Red indicates out-of-plane modes and blue indicates in-plane (transverse and longitudinal) modes.} 
\label{phonbands}
\end{figure}

The complete set of force constants are shown in \Tabref{table_ph} and the phonon dispersion relation, solving \Eqref{eigen}, is shown in \Figref{phonbands}. The solid lines represent the phonon bands of graphene on SiC and the dashed lines the dispersion of the out of plane modes of unsupported graphene. The phonon band dispersion of unsupported graphene agrees well with our published DFT based calculation. \cite{Mazzola:2017}     


\subsection{Deformation potential}

The deformation potential in the EPC matrix element $g^{\nu}(n\mathbf k,n'\mathbf {k'})$  is calculated according to the Rigid Ion Approximation (RIA), displacing a spherically symmetric screened one-electron atomic model potential\\ $V(r) = -V_0 e^{-(r/r_o)^2}$ . Then the basis orbital EPC matrix elements take the form 
\begin{eqnarray}
\langle \phi_i^I | \partial V/\partial X^{I''}_j|\phi_{i'}^{I'}\rangle =
- \langle \phi_i^I | \partial V/\partial x^{I''}_j|\phi_{i'}^{I'}\rangle &=& \nonumber \\ 
\frac{2 V_o}{r_o^2}\langle \phi_i^I |x^{I''}_j e^{-(r^{I''}/r_o)^2}|\phi_{i'}^{I'}\rangle \ . 
\label{eq:orb_matrix}
\end{eqnarray}
$X_j^I$ and $x_j^I$ denotes the Cartesian atomic displacement coordinates and electron coordinates relative the equilibrium atomic position $\mathbf R_I$, respectively. 

The parameters, strength  $V_0$ and the screening length $r_o$ are set to fit both an experimentally observed linewidth and a $first$ $principles$ calculation of EPC matrix elements. The experimental linewidth refers to the measured linewidth of the $\sigma_o$ band 200 meV below the top of the $\sigma$ bands \cite{Mazzola:2017}, and to the $first$ $principles$ calculation of the quantity
\begin{eqnarray}
 \sum_{\nu=LO,TO}|\langle \sigma_o,\mathbf k|\delta V^\nu_{\mathbf q = - \mathbf k}|\sigma_o, \mathbf o \rangle|^2 ,
\label{eq:EPC_matrix}
\end{eqnarray}
which varies weakly over the square area: - 0.1 au $\leq$ k$_x$,k$_y$ $\leq$ +0.1 au. \cite{Mazzola:2017}

\section{Linewidth - Calculations and experiment}

ARPES is a powerful tool to investigate the many-body nature of solid-state systems \cite{GaKiWe.05.Determiningelectronphonon,MaPoMi.14.Disentanglingphononand}. Indeed, it gives a direct measure of the spectral function of a material, which intrinsically contains information on the real and imaginary parts of a self energy $\Sigma$.  $\Sigma$ describes the many-body interactions, among which the most significant contributions typically come from electron phonon coupling (EPC), electron impurity scattering (EIS) and electron electron scattering (EES). For these contributions we can write $\Sigma=\Sigma^{EPC}+\Sigma^{EIS}+\Sigma^{EES}$. In addition to this, the linewidth of the ARPES spectra is closely related to the imaginary part of $\Sigma$ and it is therefore necessarily affected by all these contributions \cite{Kirkegaard:2005}.  Whilst the ARPES linewidth intrinsically contains contributions from all relevant many-body interactions, EPC is commonly responsible for abrupt changes in the linewidth. Furthermore, such abrupt changes will occur on an energy-scale corresponding to the energy of the relevant phonon mode(s). These factors generally allow the EPC contribution to the linewidth to be disentangled from EIS and EES \cite{Kirkegaard:2005,MaPoMi.14.Disentanglingphononand}.

In this section we compare the linewidth extracted form ARPES measurements with our corresponding tight-binding calculated linewidths, due to EPC. We will focus on the linewidth of the sigma bands $\sigma_o$ and $\sigma_i$ and the $\pi$ band in the high symmetry directions of the Brillouin zone. We aim at understanding which phonon modes are most important in assisting the electron scattering and to judge the relative importance of interband and intraband scattering.

\subsection{$\sigma$ bands}

\begin{figure}
\hspace{-0.2cm}
\includegraphics[width=1.\columnwidth]{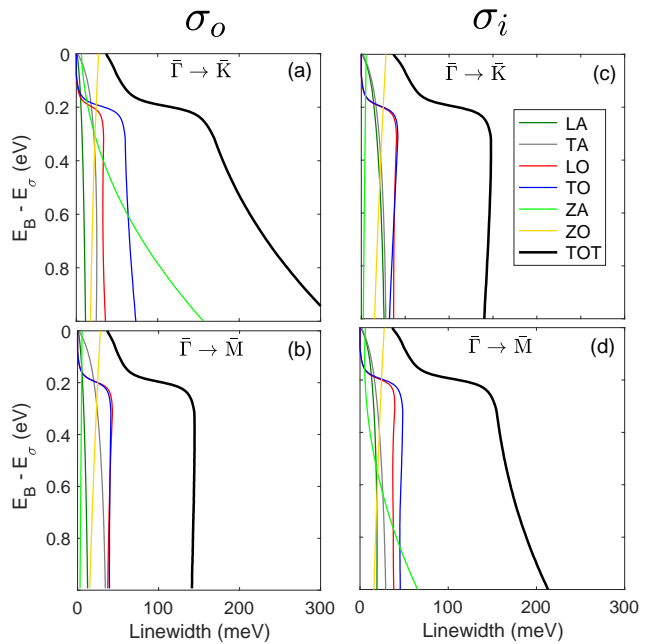}
\caption{(Color online) Calculated linewidth of the sigma bands $\sigma_o$ and $\sigma_i$ versus the binding energy of the $\sigma$ band maximum at the $\bar{\Gamma}$ point, E$_\sigma$. The black solid line represents the full linewidth and the contributions from different assisting phonon modes are shown. Results are shown for the two high symmetry directions $\bar{\Gamma} \rightarrow $ \={K} and $\bar{\Gamma} \rightarrow$ \={M}. The inset shows the color notation for the contribution to the linewidth from the different phonon modes.
}
\label{TB_sigma} 
\end{figure}

We analyze the origin of the observed kink in the $\sigma$ bands about 200 meV below the top of the $\sigma$ bands, referring to recently presented ARPES data.\cite{MaWeYa.13.Kinksinsigma, Mazzola:2017} In \Figref{TB_sigma} we show the calculated linewidth of the inner and outer $\sigma$ bands ($\sigma_i$ and $\sigma_o$) in the two high symmetry directions $\bar{\Gamma} \rightarrow$ \={K} and $\bar{\Gamma} \rightarrow $ \={M}. 

The binding energy range of about 1 eV below the top of the occupied $\sigma_o$ and $\sigma_i$ bands corresponds to a region close to the $\bar{\Gamma}$-point. The sudden increase of the calculated total linewidth ($\sigma_o$: sum of contributions from $\sigma_o\rightarrow\sigma_o$ , $\sigma_i\rightarrow\sigma_o$ and $\pi\rightarrow\sigma_o$ and $\sigma_i$: sum of contributions from $\sigma_o\rightarrow\sigma_i$ , $\sigma_i\rightarrow\sigma_i$ and $\pi\rightarrow\sigma_i$) is found in both symmetry directions at about 200 meV below the top of these sigma bands. 
%
The main contributions originate from $\sigma$ inter- and intraband scattering assisted by the two high energy optical phonon modes LO and TO. 

Analysing the linewidth of the $\sigma_o$ band in \Figref{TB_sigma}, panels (a) and (b), it is interesting to note that in the $\bar{\Gamma} \rightarrow $ \={K} direction there is an increasing  contribution from the interband scattering  $\pi\rightarrow\sigma_o$ assisted by the out-plane acoustic ZA mode for increasing binding energies.
In the direction $\bar{\Gamma} \rightarrow$ \={M} the out-of-plane ZA mode driven interband scattering \( \pi \rightarrow \sigma_o \) is of minor importance. 

The linewidth of the inner $\sigma$ band $\sigma_i$ is shown in panels (c) and (d) in \Figref{TB_sigma}. The result is reversed. The ZA mode driven $\pi \rightarrow \sigma_i$ scattering is in this case more important in the  $\bar{\Gamma} \rightarrow$ \={M} direction. The reason for this is to be found in the EPC matrix element. In the direction $\bar{\Gamma} \rightarrow$ \={K} then $\langle \sigma_o|\delta V_{ZA}|\pi\rangle$ is nearly totally symmetric while in the direction $\bar{\Gamma} \rightarrow$ \={M}, $\langle \sigma_i|\delta V_{ZA}|\pi\rangle$ is nearly totally symmetric.

We conclude that the sudden increase of the calculated full linewidth of the $\sigma_o$ and $\sigma_i$ bands at about 200 meV below the top of these bands is clear and in good agreement with experimental findings. \cite{Mazzola:2017} The sudden increase of the linewidth, $\Gamma$ = 2 Im$\Sigma^{EPC}$ , connects to a sudden change - a kink - in the observed band energy, $\varepsilon_o$ = $\varepsilon_o^0$ + Re$\Sigma^{EPC}$, where $\Sigma^{EPC}$ represents the EPC self energy. For the $\sigma_o$ band the origin stems from the $\sigma_i \rightarrow \sigma_o$ and $\sigma_o \rightarrow \sigma_o$ scattering and in the case of the $\sigma_i$ band from $\sigma_i \rightarrow \sigma_i$ and $\sigma_o \rightarrow \sigma_i$ scattering. The calculations show that the main contributions originate from assisting TO and LO phonons. 

Furthermore, we find that the linewidth of $\sigma_o$ and $\sigma_i$ bands are anisotropic in the surface Brillouin zone in the energy region investigated. The increasing contribution from the interband $\pi \rightarrow \sigma$ scattering, assisted by the ZA phonon mode, indicates that this anisotropy will be even more pronounced at greater binding energies.

\begin{figure*}
\hbox{\hspace{0.05cm}\includegraphics[width=\textwidth]{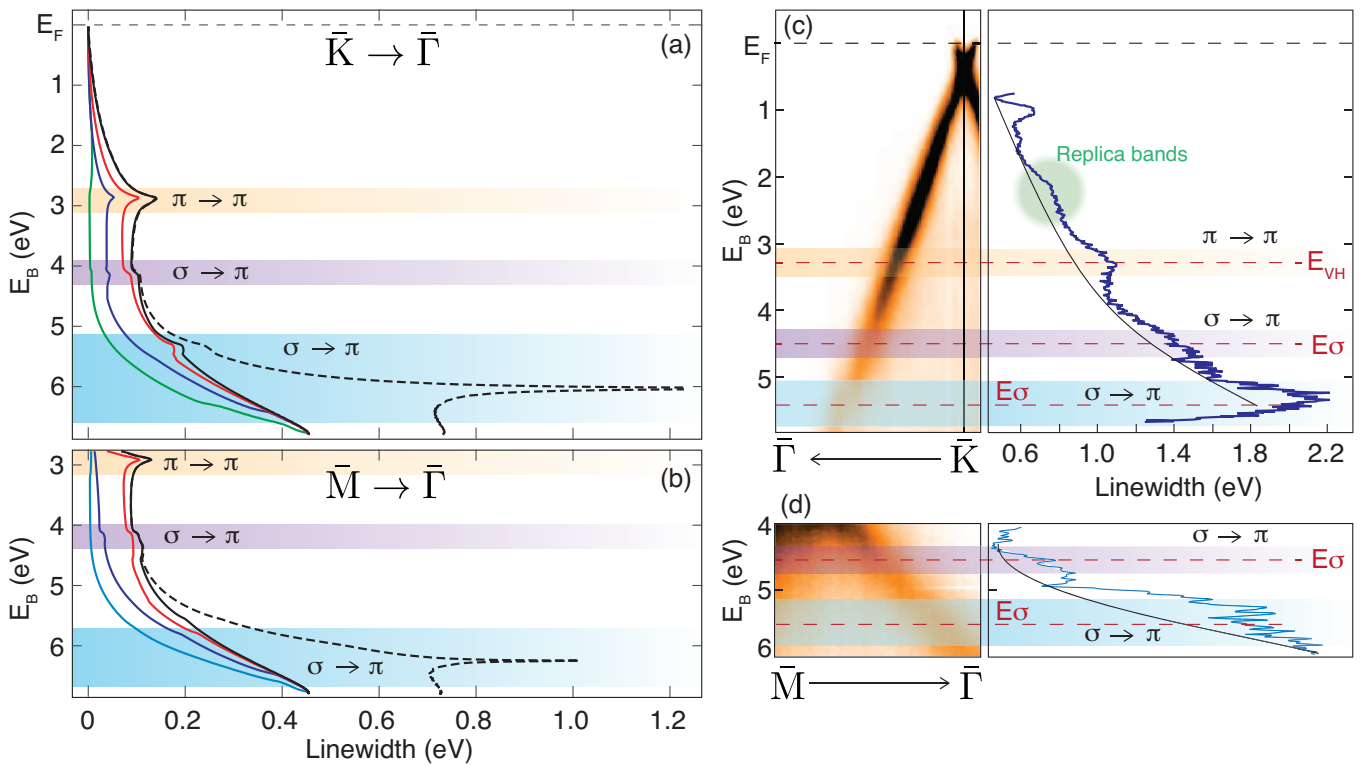}}
\caption{(Color online) Calculated and measured linewidth of the $\pi$ band of graphene supported by SiC. (a) Calculated linewidth in the  \={K}-$\bar{\Gamma}$ direction and (b) in the  \={M}-${\bar{\Gamma}}$ direction. Calculations are shown for four different vibrational energy cut-offs; 50 meV (green), 100 meV (blue), 150 meV (red) and 200 meV (black). The dashed line is the total contribution for the \textit{unsupported} graphene. (c) and (d) ARPES measurements and corresponding linewidth values. (c) ARPES data (gold-black shading) for the $\pi$ band of monolayer graphene on SiC acquired along the ${\bar{\Gamma}}$- \={K} direction of the Brillouin zone, together with the extracted linewidth (blue curve) and spectra.  To help the data visualization a baseline has been overlaid to the data. (d) ARPES data along the \={M}- $\bar{\Gamma}$ direction (gold-black shading) and the extracted linewidth. In all pictures, the yellow, purple and blue areas represent the energy ranges where the intra and inter band transitions manifest. The experimentally determined values of the $\sigma$ band maximum and VH-singularity as indicated. Green shading indicates the contributions to the linewidth given by the replica bands \cite{Nakatsuji:2010}.
}
\label{exp_theory}
\end{figure*}

\subsection{$\pi$ band}
The linewidth of the $\pi$ band has also been investigated applying our tight-binding model based calculations. In \Figref{exp_theory} we show the result of the calculation, along the high symmetry directions \={K} $\rightarrow \bar{\Gamma}$ and \={K} $\rightarrow$ \={M}. 

The results for the linewidth for the four different phonon frequency cut-offs, 50, 100, 150 and 200 meV show that the optical high frequency modes dominates the intra $\pi$ band scattering from the $\bar{M}$ point down to the top of the $\sigma$ band. 

The peak at E$_B$ $\approx$ 3 eV arises because of the increased electronic density of states due to the flat $\pi$ band in region of the $\bar{M}$ point (a.k.a. the van Hove singularity) and is discussed further below. 

Below the $\sigma$ band maximum, E$_B$ $\approx$ 4 eV, the interband scattering $\sigma_i \rightarrow \pi$ and $\sigma_o \rightarrow \pi$ becomes increasingly more important. The green line in \Figref{exp_theory}, corresponding to a phonon energy maximum of 50 meV clearly indicates that it is only the acoustic out-of-plane ZA mode which is in operation in the $\sigma_i \rightarrow \pi$ and $\sigma_o \rightarrow \pi$ scattering. As the bottom of the $\pi$ is approached, the contribution from these scattering channels dominate completely. This is explained by the increase of the phase space of the initial electron states, referring to the $\sigma_i$ and $\sigma_o$ bands and in addition also to the reduced slope of the $\pi$ band as the $\bar{\Gamma}$ point is reached.

The peak in the linewidth of the $\pi$ band of unsupported graphene (dashed lines in \Figref{exp_theory} (a) an (b)) close to the bottom of the $\pi$ band is due to a large contribution near the crossing of the $\pi$ band and the $\sigma_o$ band and the $\sigma_i$ band in the direction \={K} $\rightarrow \bar{\Gamma}$ ( \Figref{exp_theory}  (a)) and \={M} $\rightarrow \bar{\Gamma}$ ( \Figref{exp_theory}  (b)), respectively. This peak signals the instability of unsupported graphene due to the $\sim q^2$ dispersion of the ZA mode. This singularity is lifted when graphene is supported on a substrate.\cite{Amorim:2013} The finite frequency, $\sim$ 24 meV, at the $\bar{\Gamma}$-point (see \Figref{phonbands}) stabilizes the crystal structure of the graphene layer.

The reason why the peaks appear at different binding energies is again, just as for the $\sigma$ bands, to be found in the EPC matrix element. Considering the unit cell including the $A$ and the $B$ atom, we have that in the $\bar{\Gamma} \rightarrow $ \={K} direction $|\sigma_o\rangle \approx \frac{1}{\sqrt{2}}(|2p^A_x\rangle -|2p^B_x\rangle )$ (with the $x$-axis in the $A$ to $B$ direction) while in the $\bar{\Gamma} \rightarrow $ \={M} direction $|\sigma_i\rangle$ has the same form. Thus $\langle \pi|\delta V_{ZA}|\sigma_o\rangle$ will be totally symmetric in the  $\bar{\Gamma} \rightarrow $ \={K} direction, while in the $\bar{\Gamma} \rightarrow $\={M} the $\langle \pi|\delta V_{ZA}|\sigma_i\rangle$ will be totally symmetric.

ARPES measurements on monolayer graphene on SiC and their corresponding spectral linewidth are shown in Fig.\ \ref{exp_theory}~(c) and (d) for the $\pi$ band of graphene acquired along the $\bar{\Gamma}$-\={K} and $\bar{\Gamma}$-\={M} directions, respectively. Each cut at constant energy in Fig.\ \ref{exp_theory}~(c,d), a so-called momentum distribution curve (MDC), has been fitted by Lorentzian curves with inclusion of a cubic polynomial background and from the fit results, the linewidth as a function of energy is extracted. As can be seen in the figure, the linewidth shows several sudden changes occurring at E$_{B} \approx 2$;   $3.3$;   $4.5$ and $5.5$~eV. Unlike our calculation (which only includes EPC contributions to the linewidth), the ARPES measurement intrinsically includes all relevant interactions. It is therefore necessary to discuss the origin of the experimentally observed linewidth changes.

At E$_{B} \approx 2$~eV, (green area in Fig.\ \ref{exp_theory}~(b)) a change in the spectral linewidth is observed. At such an energy the graphene/SiC electronic dispersion is known to be affected by replica-bands. These bands originate from the interaction between the graphene and the substrate on which it is grown  \cite{Nakatsuji:2010}. These bands have a weak intensity and in our experimental data are difficult to see, however the lineshape of the ARPES spectra in this region indicates the presence of additional components, and we conclude that they are responsible for the linewidth change observed experimentally at this $E_B$ value.

The linewidth changes at E$_{B}\approx 3.3$~eV and E$_{B}\approx 4.5$~eV cannot be explained by replica bands or substrate interactions (Fig.\ \ref{exp_theory}(c-d), orange and purple areas, respectively): At $E_{B}\approx 3.3$~eV the e-DOS suddenly increases, due to the electron accumulation at the Van Hove (VH) singularity \cite{VanHove:1953}. The VH-singularity constitutes a local maximum of the $\pi$-band at the M point of the BZ, as indicated in Fig.\ \ref{ebands}. Therefore, the VH-singularity creates an increase in the e-DOS, and hence the probability of phonon mediated refilling of the photo-hole is dramatically increased, in good agreement with our  calculations (for example, Fig.\ \ref{exp_theory} (a). 

At $E_{B}\approx 4.5$~eV, a similar change in the measured linewidth is also seen. This also occurs at an energy where the e-DOS is dramatically increased, but in this case it is because of the maximum of the $\sigma$-band. Again, because the e-DOS shows a strong increase, the probability of phonon-mediated refilling dramatically increased and hence the lifetime of a photo-hole is reduced. In agreement with our tight-binding calculation, this is observed as an increase in the linewidth due to EPC. 

At $E_{B}\approx 5.5$~eV, the ARPES linewidth has become very broad, and also appears to show a peak. It is difficult to unambiguously disentangle the EPC contribution to the linewidth since EES may also play a significant role here. Also, the dramatically increased ARPES linewidth hinders accurate analysis. However, it is interesting to note that the tight-binding calculation (which only includes EPC contributions to the linewidth) predicts that the linewidth will dramatically increase in this $E_B$ range, hence it seems feasible that the large measured linewidth is at least partially due to increased EPC. Unlike the previous cases where the increase in EPC was primarily due to an increase in the e-DOS, in this case, it is the crossing of the $\pi$-band with $\sigma_i$ and $\sigma_o$ which dramatically increases the efficiency of EPC by allowing phonon modes with little energy and momentum (i.e. acoustic modes) to make a large contribution.

It should also be noted that the experimentally reported changes in the spectral linewidth are small and, purely from the experiment, we cannot exclude \textit{a priori} that there might be other contributions to such linewidth changes; however, the good agreement between the experiment and the tight-binding calculation (which predicts the same linewidth changes) significantly strengthens the validity of our interpretations.\\ 


\section{Summary and Conclusions}

We present a theoretical investigation of the electron-phonon interaction in pristine graphene and compare with experimental ARPES data. The interaction is found to be considerably stronger in the $\sigma$ band than in the $\pi$ band.

The theoretical linewidth analysis of the two uppermost occupied $\sigma$ bands in the region of the $\bar{\Gamma}$-point supports the picture that the scattering is primarily driven by the high energy optical phonon modes LO and TO. 
The calculations also reveal a strong anisotropy of these $\sigma$ bands in the surface Brillouin zone. In the $\bar{\Gamma} \rightarrow $  \={K} direction, the interband scattering $\pi \rightarrow \sigma$, driven by the out-of-plane phonon mode ZA, dominates in most of the energy region where the $\sigma$ and $\pi$ bands overlap.

The calculated linewidth of the $\pi$ band is compared in detail along the \={K} $\rightarrow $  $\bar{\Gamma}$ and \={M} $\rightarrow $  $\bar{\Gamma}$ directions with ARPES data. The main features are reproduced by the calculations. In the energy regions where the $\sigma$ and $\pi$ overlap the linewidth is found to be nearly isotropic in the surface Brillouin zone. Also for the $\pi$ band, the interband scattering, now $\sigma \rightarrow \pi$, dominates and the acoustic ZA mode is most important.

We show that in order to understand the variation of the linewidth it is not enough to only consider the density of state effects (for example the van Hove singularities) - it is also important to consider the symmetry of the EPC matrix element. The latter is of central importance in some regions of the BZ (for example, at the $\sigma$-band maximum). We also demonstrate that when taking the graphene-substrate coupling into account, the lattice instability of unsupported graphene caused by the acoustic ZA vibrational mode is removed and the sharply peaked $\pi$-band linewidth increase is reduced such that it is in better agreement with the experimental data.

\acknowledgments{}
This work was partly supported by the Research Council of Norway through its Centres of Excellence funding scheme, project number 262633, ``QuSpin'', and through the Fripro program, project number 250985 ``FunTopoMat''.
The linewidth calculations were performed on resources at Chalmers Center for Computational Science and Engineering (C3SE) provided by the Swedish National Infrastructure for Computing (SNIC).
TF acknowledges Grant.~FIS2017-83780-P from the Spanish Ministerio de Econom{\'i}a y Competitividad.


\bibliographystyle{apsrev-title}
\bibliography{graphene-clean.bib}

\end{document}